\documentclass[twocolumn,superscriptaddress,
showpacs,preprintnumbers,amsmath,amssymb]{revtex4}
\usepackage{amssymb,amsmath,amsthm}
\usepackage[all]{xy}
\usepackage{multirow}
\usepackage{graphicx}

\theoremstyle{definition}

\newcommand{\ket}[1]{{\left| #1 \right\rangle}}

\newcommand{\T}{\mbox{$\mathrm{tr}$}}

\begin{document}
%%%%%%%%%%%%%%%%%%%%%%%%%%%%%%%%%%%%%%%%%%%%%%%%%%%%%%%%%%%%%%%%%%%%%%%%%%
%                                                                        %
%                                 Title                                  %
%                                                                        %
%%%%%%%%%%%%%%%%%%%%%%%%%%%%%%%%%%%%%%%%%%%%%%%%%%%%%%%%%%%%%%%%%%%%%%%%%%
\title{Security problem on arbitrated quantum signature schemes }

\author{Jeong Woon Choi}\email{jw_choi@etri.re.kr, cju@snu.ac.kr}
\affiliation{
Cryptography Research Team,
Electronics and Telecommunications Research Institute, Daejeon 305-700, Korea }
\author{Ku-Young Chang}%\email{jang1090@etri.re.kr}
\affiliation{
Cryptography Research Team,
Electronics and Telecommunications Research Institute, Daejeon 305-700, Korea }
\author{Dowon Hong}%\email{dwhong@etri.re.kr}
\affiliation{
Cryptography Research Team,
Electronics and Telecommunications Research Institute, Daejeon 305-700, Korea }

\date{\today}

%%%%%%%%%%%%%%%%%%%%%%%%%%%%%%%%%%%%%%%%%%%%%%%%%%%%%%%%%%%%%%%%%%%%%%%%%%
%                                                                        %
%                              Abstract                                  %
%                                                                        %
%%%%%%%%%%%%%%%%%%%%%%%%%%%%%%%%%%%%%%%%%%%%%%%%%%%%%%%%%%%%%%%%%%%%%%%%%%
\begin{abstract}
Until now, there have been developed many arbitrated quantum signature
schemes implemented with a help of a trusted third party. In order
to guarantee the unconditional security, most of them take
advantage of the optimal quantum one-time encryption method based on
Pauli operators. However, we in this paper point out that
the previous schemes only provides a security against total break
and actually show that there exists a simple existential forgery attack to validly modify
the transmitted pair of message and signature.
In addition, we also provide a simple method to recover the security against the proposed attack.
\end{abstract}

\pacs{ %89.70.+c, %Information theory and communication theory
%03.65.Ta, % Foundations of quantum mechanics; measurement theory
%03.65.Ud, % Entanglement and quantum non-locality
03.67.Dd, % Quantum Cryptography
%03.67.-a, % Quantum information
03.67.Hk % Quantum communication
%03.67.Mn  % Entanglement production, characterization and manipulation
}

\keywords{arbitrated quantum signature schemes, quantum one-time
encryption, Pauli operators, existential forgery}

\maketitle

%%%%%%%%%%%%%%%%%%%%%%%%%%%%%%%%%%%%%%%%%%%%%%%%%%%%%%%%%%%%%%%%%%%%%%
%%%                                                                %%%
%%%                         Introduction                           %%%
%%%                                                                %%%
%%%%%%%%%%%%%%%%%%%%%%%%%%%%%%%%%%%%%%%%%%%%%%%%%%%%%%%%%%%%%%%%%%%%%%
\section{Introduction}

Digital signature schemes make possible a variety of cryptographic applications
such as authentication of message origin, data integrity, non-repudiation and so on.
However, the advent of quantum computing algorithm has fatally weakened the security of the public-key cryptosystems especially based on the discrete-log
or factoring problems and thus also clearly the security of digital signature schemes.
Therefore, in order to guarantee the security of signature schemes even against the unlimited computational power of attackers,
it is so meaningful to develop quantum analogues of digital signature schemes.

In 2001, Gottesman and Chuang~\cite{GC} provided a quantum signing and verifying method for digital messages by using quantum one-way function and quantum swapping test~\cite{BCWW}.
Since then, there have been various efforts to extend the domain of signable messages to arbitrary known and unknown quantum messages.
For example, Zeng and Keitel~\cite{ZK} proposed an arbitrated quantum signature (AQS) scheme based on the correlation of Green-Horne-Zeilinger (GHZ) states~\cite{GHZ}
and quantum symmetric encryption method~\cite{BR,AMTW},
and many variations~\cite{CL,Z,LCL,ZQ,L} of this method have been developed until recently.
Note that the security of these AQS schemes depends on the fact that the secret key $K_{AT}$ shared between a signer, Alice, and an arbitrator, Trent,
is kept secretly from attackers including a verifier, Bob.
That is, they insist that it is impossible for an attacker to forge a signature of Alice because of the ignorance of the secret key.

However, we show in this paper that there exists a forgery attack to make a new valid signature pair from the original pair of the transmitted message and signature
in almost all AQS schemes using quantum one-time encryption,
even though a dishonest party does not have any information of the secret key $K_{AT}$.
We would like to emphasize that when introducing the quantum encryption scheme to construct a quantum signature scheme,
something more than the secrecy of the secret key should be considered to prove the security of the quantum signature scheme.
Actually, the AQS schemes are derived from classical arbitrated signature schemes.
The reason why those classical schemes are secure although they are neither provably nor unconditionally secure,
more concretely, the reason why they guarantee the data integrity of the transmitted message is that
they combine the hash function with the encryption in order to intertwine the message bits and so detect any kind of tampering.
In addition, we also provide a simple method to enable the AQS protocols to circumvent our existential forgery attack.

%%%%%%%%%%%%%%%%%%%%%%%%%%%%%%%%%%%%%%%%%%%%%%%%%%%%%%%%%%%%%%%%%%%%%%
%%%                                                                %%%
%%%                          Preliminaries                         %%%
%%%                                                                %%%
%%%%%%%%%%%%%%%%%%%%%%%%%%%%%%%%%%%%%%%%%%%%%%%%%%%%%%%%%%%%%%%%%%%%%%
%\section{Preliminaries}
%1.attacker's aim
%2.quantum one-time encryption
%3.swapping test for verification

%%%%%%%%%%%%%%%%%%%%%%%%%%%%%%%%%%%%%%%%%%%%%%%%%%%%%%%%%%%%%%%%%%%%%%
%%%                                                                %%%
%%%                 Review of arbitrated QS protocols              %%%
%%%                                                                %%%
%%%%%%%%%%%%%%%%%%%%%%%%%%%%%%%%%%%%%%%%%%%%%%%%%%%%%%%%%%%%%%%%%%%%%%

\section{A brief review of the ZK protocol}
In this section we briefly review the ZK protocol proposed by Zeng and Keitel~\cite{ZK}
instead of introducing all AQS protocols.
The ZK protocol has actually the most complicate structure compared to other variations of the protocol.
Our attack method described in the next section~\ref{Sec:Security} to forge the quantum signature in the ZK protocol
will be more easily applied to other AQS protocols in~\cite{CL,Z,LCL,ZQ,L}.

The ZK protocol consists of three phases: Initializing phase, Signing phase, Verifying phase.
\subsection{Initial phase}
\begin{itemize}
\item[A1:] Alice and Bob each share secret key strings $K_{AT}$ and $K_{BT}$ with Trent by using a practical quantum key distribution protocol~\cite{BB,LC02,MHS,H01,U,SRS}.
\item[A2:] Trent generates the GHZ triplet states, $$\frac{1}{\sqrt{2}}(\ket{000}+\ket{111}).$$
               For each GHZ states, Trent holds one particle of it for himself and distributes each of the remaining two particles to Alice and Bob.
\end{itemize}
\subsection{Signing phase}
\begin{itemize}
\item[B1:] Alice creates a message string of $n$ qubits, $\ket{P}=\otimes_{i=1}^n \ket{P_i}$ and two copies of it.
\item[B2:] Alice obtains $\ket{R}$ by applying to a random rotation $R_{K_{AT}}$ to one copy of $\ket{P}$ according the secret key $K_{AT}$ as follows
                \begin{eqnarray*}
                \ket{R}&=&R_{K_{AT}}\ket{P_i}\\
                       &=&\otimes_{i=1}^n \sigma_x^{K_{AT_i}}\sigma_z^{K_{AT_i}+1}\ket{P_i}.
                \end{eqnarray*}
\item[B3:] By performing Bell measurement on each particle of the other copy of $\ket{P}$ and the GHZ state, Alice also obtains $2n$-bit string $\mathcal{M}_A$.
                These measurement outcomes are given by the following relation:
\begin{eqnarray}\label{GHZ relation}
            &&\ket{P_i}\otimes\ket{GHZ}\nonumber\\
            &=&(a_i\ket{0}+b_i\ket{1})\otimes\frac{1}{\sqrt{2}}(\ket{000}+\ket{111})\\
            &=&\frac{1}{2\sqrt{2}}\big[\ket{\Phi^+}\{\ket{+}(a_i\ket{0}+b_i\ket{1})+\ket{-}(a_i\ket{0}-b_i\ket{1})\} \nonumber \\
            & &+\ket{\Phi^-}\{\ket{+}(a_i\ket{0}-b_i\ket{1})+\ket{-}(a_i\ket{0}+b_i\ket{1})\}\nonumber \\
            & &+\ket{\Psi^+}\{\ket{+}(b_i\ket{0}+a_i\ket{1})+\ket{-}(b_i\ket{0}-a_i\ket{1})\} \nonumber\\
            & &-\ket{\Psi^-}\{\ket{+}(b_i\ket{0}-a_i\ket{1})+\ket{-}(b_i\ket{0}+a_i\ket{1})\}\big], \nonumber
\end{eqnarray}
where $\ket{\pm}=\frac{\ket{0}\pm\ket{1}}{\sqrt{2}}$ and four Bell states are given as $\ket{\Phi^{\pm}}=\frac{\ket{00}\pm\ket{11}}{\sqrt{2}}$ and
$\ket{\Phi^{\pm}}=\frac{\ket{01}\pm\ket{10}}{\sqrt{2}}$.

\item[B4:] Alice makes a signature
                $$\ket{S(P)}=E_{K_{AT}}(\mathcal{M}_A,\ket{R})$$
                by applying a quantum one-time encryption $E_{K_{AT}}$ to $\mathcal{M}_A$ and $\ket{R}$,
                where $E_{K_{AT}}=\otimes_{i=1}^n \sigma_x^{K_{AT_{2i}}}\sigma_z^{K_{AT_{2i+1}}}$.
\item[B5:] Alice sends the quantum message $\ket{P}$ and its corresponding quantum signature $\ket{S(P)}$ to Bob.
\end{itemize}

\subsection{Verifying phase}
\begin{itemize}
\item[C1:] By performing the $X$-basis measurement on his particles of the GHZ states, Bob obtains $n$-bit string $\mathcal{M}_B$
                and sends Trent the quantum states $\ket{Y_B}=E_{K_{BT}}(\mathcal{M}_B,\ket{P},\ket{S(P)})$ obtained by encrypting $\mathcal{M}_B$ and $\ket{P},\ket{S}$ received from Alice
                according to $K_{BT}$.
\item[C2:] Trent decrypts $\ket{Y_B}$ with $K_{AT}$ and $K_{BT}$ and then obtains $\ket{P}$, $\mathcal{M}_A$, $\mathcal{M}_B$, and $\ket{R}$.
                Then, he tests if the decryption result satisfies $R_{K_{AT}}\ket{P}=\ket{R}$. Of course, if the quantum message $\ket{P}$ is known, then
                an orthogonal measurement will be used for the equality test, and otherwise, the swapping test will be adopted.
\item[C3:] With the test result, Trent sends back to Bob $\ket{P}$, $\mathcal{M}_A$, $\mathcal{M}_B$, and his particles of the GHZ states.
\item[C4:] If the test result is TRUE, Bob adequately applies the Pauli operators according to the GHZ relation in Eq.~(\ref{GHZ relation}) in order to
                recover the original quantum message from the particles of GHZ state received from Trent.
                If there were no dishonest actions during the previous procedures, then the recovered quantum message will be same to the original message
                and Bob verifies the equality by the same way to the test by Trent.
\end{itemize}

%%%%%%%%%%%%%%%%%%%%%%%%%%%%%%%%%%%%%%%%%%%%%%%%%%%%%%%%%%%%%%%%%%%%%%
%%%                                                                %%%
%%%                 Review of arbitrated QS protocols              %%%
%%%                                                                %%%
%%%%%%%%%%%%%%%%%%%%%%%%%%%%%%%%%%%%%%%%%%%%%%%%%%%%%%%%%%%%%%%%%%%%%%
\section{Security analysis for Arbitrated Quantum Signature protocols}\label{Sec:Security}
In this section, we focus on analyzing the security of the ZK protocol.
Note that their security flaws are due to the usage of the quantum one-time encryption method based on Pauli operators.
While quantum encryption is for hiding quantum information securely,
quantum signature schemes must have additional functionalities such as the tamper-proof of the signed quantum data.

The present AQS protocols are concentrated on proving the security only against the total break attack, that is,
whether or not the secret key can be distilled by attackers from the transmitted pair of quantum message and signature.
Unfortunately, we here show that there exists an existential attack which enables a dishonest party to modify the quantum message and signature
to a new valid pair, even though the attacker has no information of the secret key.
Of course, if the quantum message is publicly known, then this attack will be beyond only the existential attack.

\subsection{The details of the new attack method against AQS protocols }
We here introduce an attack method to change the pair of quantum message and signature validly without the knowledge of the secret key.
In this section, we only deal with the case that the attacker is the verifier, Bob.
However, any eavesdroppers not participated in the protocol can also perform the same attack.

The core of our attack method is to show how Bob does change the signature pair $\left(\ket{P},\ket{S(P)}\right)$
to a new one $\left(\ket{P'},\ket{S(P')}\right)$ without the knowledge of $K_{AT}$.
This is accomplished by using the anti-commutativity of non-trivial Pauli operators as follows:
\begin{eqnarray*}
\sigma_x\sigma_z&=&-\sigma_z\sigma_x\\
\sigma_z\sigma_y&=&-\sigma_y\sigma_z\\
\sigma_y\sigma_x&=&-\sigma_x\sigma_y
\end{eqnarray*}
Since the global phase has no physical meaning in quantum mechanics,
the above relation after all implies that
any changes by Pauli operators are commutative with the quantum signing process consisting of the quantum one-time encryption based only on Pauli operators.
Therefore, we can regard any Pauli changes after the signing process as
the Pauli changes before the signing process,
that is, the quantum signatures generated in the AQS protocols can be always forged to quantum signatures of the quantum message changed by the Pauli operators.

We have two types of approaches to show the security violations of the ZK protocol according to whether Trent makes use of the classical information $\mathcal{M}_A$ crucially in the verifying phase.

First, if $\mathcal{M}_A$ is not involved in the test of the validity of quantum signature as shown in the original paper, then the attack is more simple.
When Bob receives the pair of quantum message and signature from Alice,
he selects a Pauli operator $U$ and applies it to the signature pair.
This selection would be implemented intentionally if the quantum message is publicly known, and randomly if it is unknown.
This type of attack is always successful, because of the following relation:
\begin{eqnarray}\label{attack1}
\left(U\ket{P},U\ket{S(P)}\right)&=&\left(\ket{P'},U E_{K_{AT}}(\mathcal{M}_A,R_{K_{AT}}\ket{P})\right) \nonumber\\
                                 &=& \left(\ket{P'}, E_{K_{AT}}(\mathcal{M}_A,\alpha U R_{K_{AT}}\ket{P})\right) \nonumber\\
                                 &=& \left(\ket{P'}, E_{K_{AT}}(\mathcal{M}_A,\beta R_{K_{AT}}U\ket{P})\right) \nonumber\\
                                 &\equiv& \left(\ket{P'}, E_{K_{AT}}(\mathcal{M}_A, R_{K_{AT}}\ket{P'})\right) \nonumber\\
                                 &=& \left(\ket{P'}, \ket{S(P')}\right),
\end{eqnarray}
where $\alpha$ and $\beta$ are $\pm 1$ and $\equiv$ means the equality up to the global phase.

Otherwise, suppose that
Trent recovers $\ket{P}$ from his particle of the GHZ state by using the bit information of $\mathcal{M}_A$ and $\mathcal{M}_B$
and then check if the result is equal to the decrypted result of $R\ket{P}$.
Then, dishonest Bob must find an attack
to modify both of $\mathcal{M}_A$ and $R\ket{P}$ adequately in order not to be detected in the verification by Trent.
We would like to emphasize again that the important thing for an attacker is not what the transmitted information pairs are but
how to use the relation they have.
There also exists a deterministic way to modify $\mathcal{M}_A$ to $\mathcal{M'}_A$
rightly according to what Bob wants to change the transmitted quantum message to.

For example, suppose that Bob tries to change the quantum message from $\ket{P}$ to $\sigma_x\ket{P}$
and thus apply $\sigma_x$ to the quantum part of $\ket{S(P)}$ as in Eq.~(\ref{attack1}).
Let's look at the Eq.~(\ref{GHZ relation}) carefully.
When Alice and Bob have completed Bell measurement and $X$-basis measurement, respectively,
the measurement outcome of Alice, $\mathcal{M}_A$, and the final state of Trent's particle of the GHZ state, $\ket{Q}$, have
the specific relation, regardless of Bob's measurement result, $\mathcal{M}_B$.
In Eq.~(\ref{GHZ relation}), the exchange of $\mathcal{M}_A$ such as
$\ket{\Phi^+}\leftrightarrow\ket{\Psi^+}$ and $\ket{\Phi^-}\leftrightarrow\ket{\Psi^-}$
exactly corresponds to the exchange of $\ket{Q}$ by $\sigma_x$, regardless of $\mathcal{M}_B$.
If Bob forges $\mathcal{M}_A$ to $\mathcal{M'}_A$ by the above exchange relation, then Trent will try to recover the original $\ket{P}$
by applying $\sigma_x$ to $\ket{Q}$.
However, this behavior reversely changes the original $\ket{P}$ to $\sigma_x\ket{P}$ and Bob's attack will success.
Note that the forgery of classical messages $\mathcal{M}_A$ is always possible,
because the ZK protocol and its variations adopt the quantum or classical bitwise one-time pad for them and
the bit flip of encrypted message is exactly same to the encryption of the bit flipped message.

\subsection{The security recovery against the new attack }
As described in the previous section, the present AQS schemes are cracked by our existential forgery attack.
This is because all quantum operations used for random rotation and one-time encryption are only Pauli operators
which commute or anti-commute with each other.
For a quantum message $\ket{P}$, the corresponding quantum signature will be in the form of $ER\ket{P}$, where
$R$ and $E$ represent a random rotation and an one-time encryption, respectively.
If Bob performs a forgery attack with a quantum operation $Q$, then the quantum signature turns to $ER\ket{P}$
and finally becomes $R^{\dagger}E^{\dagger}QER\ket{P}$ by Trent's decryption.
Therefore, it is necessary for an attacker without knowledge of $R$ and $E$ to set $Q$ as a quantum operation which commutes with both of $E$ and $R$.
This is the main idea of our proposed attack method.
Fortunately, this also means that if it is possible to prevent an attack from finding such a quantum operation,
then we can make the AQS protocols secure.

We here need to remind the following necessary and sufficient condition for the optimal quantum one-time encryption in~\cite{BR}.
\begin{itemize}
\item[] {\it A set $\{p_k,U_k\}_{k\in K}$ where $K$ consists of $2n$-bit secret strings is a quantum encryption set
if and only if the unitary operator elements form an orthonormal basis, and they are all equally likely.}
\end{itemize}
Consider a set of unitary operators which are given in the form of $U\{I, \sigma_x, \sigma_y, \sigma_z\} V$ for arbitrary unitary operator $U$ and $V$.
This is absolutely an optimal quantum one-time encryption, because for any Pauli operator $P$ and $P'$,
$UPV$ and $UP'V$ have the following Hilbert-Schmidt inner product~\cite{HJ},
\begin{eqnarray}
\T((UPV)^{\dagger}UP'V)=\T(P^{\dagger}P')=2\delta_{P,P'}.
\end{eqnarray}
We call it the $(U,V)$-type quantum encryption.
In order to include the non-commutative property in the quantum signing process,
we simply take advantage of the random rotation based on Pauli operators and the $(I,H)$-type quantum encryption where $H$ is the Hadamard operator defined by $H=\frac{1}{\sqrt{2}}\left(%
\begin{array}{cc}
  1 & 1 \\
  1 & -1 \\
\end{array}%
\right).$
Then the finally decrypted quantum signature will be in the form of $UHVQVHU\ket{P}$, where
$U$ and $V$ are unknown Pauli operators for a random rotation and an one-time encryption, respectively.
If Bob takes $Q$ a Pauli operator, he is able to remove $V$ without the knowledge of $V$ as shown in our attack method.
However, the Hadamard operator $H$ will prevent the spread of our attack method, because there is no non-trivial quantum operator to commute with
both of $H$ and a nontrivial Pauli operator.
Of course, if we take the $(U,V)$-type quantum encryption randomly and securely instead of using a publicly opened and fixed type of quantum one-time encryption,
it is trivial that the security of AQS protocols is increased, even though secure key bits are additionally consumed.

\subsection{Other security issues}
In this section, we introduce other security problems which should be additionally considered in most present quantum signature schemes.
\begin{itemize}
\item {\it They deal with quantum signature schemes bit-wisely or quantum bit-wisely.}
\end{itemize}
      This means that it is always possible for an attacker to permute the order of the message and signature pair as he wants.
      For example, although the attacker does not know the contents of the message,
      many messages such as official and financial documents usually have a specific form
      and thus the attacker can modify the original message by permuting the important data including date, time, the amount of money and so on.
\begin{itemize}
\item {\it Any symmetric states can pass the quantum state equality test.}
\end{itemize}
      Regardless of whether the quantum signature schemes depends on quantum one-way functions or quantum encryptions by the shared secret key string,
      in order to confirm the validity of quantum signature,
      it is necessary to test the equality of the final quantum states generated by the different routes and algorithms of the signer, the verifier, or the arbitrator.
      Up to date, the quantum swapping test proposed in~\cite{BCWW} is the only way used for achieving the purpose.
      However as noted in~\cite{GC}, any symmetric states can pass the equality test and thus
      this fact can be used maliciously for Alice's disavowal.

%%%%%%%%%%%%%%%%%%%%%%%%%%%%%%%%%%%%%%%%%%%%%%%%%%%%%%%%%%%%%%%%%%%%%%
%%%                                                                %%%
%%%                          Conclusion                            %%%
%%%                                                                %%%
%%%%%%%%%%%%%%%%%%%%%%%%%%%%%%%%%%%%%%%%%%%%%%%%%%%%%%%%%%%%%%%%%%%%%%

\section{Conclusion}
In this paper, we have pointed out that most of present AQS protocols
can be cracked by a specified existential attack.
This is due to the careless usage of quantum one-time encryption based on Pauli operators.
Generally, an encryption method is very useful for hiding a data and validating the origin of the data.
However, in order for it to satisfy the tamper-proof property in a signature scheme, it requires more complicate structure beyond the bitwise one-time encryption
as in classical cryptography hash functions are combined with the encryption to interlace the transmitted message bits and detect any modifications of them.

In addition, to overcome the weakness of the AQS schemes against our existential forgery attack,
we also proposed a method to detect the forgery attack with a non-negligible probability for each bit or quantum bit
by adding the non-commutative property on the signing process by using the extended class of quantum encryption.

\section*{Acknowledgements}
This work was supported by the R\&D program of ETRI, Korea
[Development of Privacy Enhancing Cryptography on Ubiquitous Computing Environment].

%%%%%%%%%%%%%%%%%%%%%%%%%%%%%%%%%%%%%%%%%%%%%%%%%%%%%%%%%%%%%%%%%%%%%%%%

\end{document}